\documentclass{acm_proc_article-sp}

\makeatletter
\let\@copyrightspace\relax
\makeatother
\begin{document}

\title{Examining Motivations behind Paper Usage in Academia}
%
%
%
%
%

%
\author{
%
%
\alignauthor
Joey Chakraborty\\
       \email{joey@chakraborty.ca}
}


\maketitle
\begin{abstract}
We carried out a qualitative study to identify the ``missing pieces" in current computing devices and technologies that are preventing people from eliminating paper from their lives. 

Most of the existing literature has looked into the work practices of businesses, while a few have researched how high school and college students and teaching assistants at universities work with paper. We were specifically interested in analyzing paper use for people in the research side of academia, and seeing how our results compare to existing work. We recruited and interviewed participants from academia to understand what kind of tasks they use paper for, what kind of tasks they use computing devices for and what motivates them to use these two media. 

We found that, despite having access to at least one personal computing device, the participants preferred to work with paper in many situations. This appears to be attributed to certain intrinsic qualities that paper has, such as open format, easy navigation, readability, and the affordances these qualities provide. In order to eventually replace paper with devices, designers of new technology will have to successfully emulate these qualities.

\end{abstract}

\section{Introduction}
A wide array of personal computing devices is available in the market today, each more portable and powerful than the next: smart phones, tablets, netbooks/laptops, eReaders, tablet PCs, etc. These devices offer a wide range of functionality and often become indispensable tools for the people that use them.

In spite of the widespread proliferation of these devices in our society, we still seem to heavily rely on paper, almost on a daily basis. To better understand why paper usage is so entrenched in our lives, it's important to identify the attributes of paper that lend to its pervasive quality.

\subsection{Motivation}
Paper has been used by human beings for well over a millennium. Even today, where most people have access to at least one computing device, if not more, paper use is still very much present. 

Other than being a commodity that has an environmental impact, paper can be damaged easily thereby losing all information on it. Tracking changes and updating paper documents can be cumbersome and leaves a lot of room for error. Also, paper is usually a one-time use item. Once it has been printed, drawn, or written on and has served its purpose, a piece of paper is typically disposed of. On the other hand, one might argue that paper is easy to carry, easy to find and inexpensive.

Most computing devices today include software that let us create, edit, view, share documents and more. In addition, information on these devices is easier to find, update and back-up. Last but not least, these devices can be used again and again over the lifetime its user, to consume and distribute information. However, battery life, size and cost can often be an issue when it comes to adopting these devices.

Given the apparent advantages of devices over paper, one might think that the latter would not be used as much as it is in this day and age. Upon observing work practices of people around us, it is clear that paper is still the tool of choice. What is not clear is why that is the case. In this study, we attempt to learn more about the qualities that make paper so desirable to work with. The better we understand these qualities, the better job we can do in incorporating them into devices, which in turn would move us towards a paperless world.

In order to consider diverse opinions, it was important to find participants who use devices and paper frequently and heavily. Academia happens to be a field that deals with a vast amount and variety of information that gets distributed and shared on a regular basis. It also happens to be an area where a lot of people own multiple devices, some of which are often brand new and top of the line. Also, academia seems to particularly experience heavy paper usage. Given these attributes of the field, people in academia are an ideal choice of participants for this study.

\subsection{Related Work}

The idea of ``going paperless" is not a new one. A fair bit of research has been done in this area to see if it is indeed possible to completely digitize work in the office, school and other settings and how that might be done.

One of the more prominent publications in this domain is a book written by Sellen and Harper \cite{sellen:myth}. Sellen et al. trace the first idea of the existence of a paperless office to the 1800s with Samuel Morse's idea of sending message electronically via the telegraph system. After examining paper usage trends around the world through the decades, they conclude that paper usage has in fact grown with the advancement of technology. Sellen et al. go on to say that paper offers certain affordances that are better suited to the way that people work, than technologies do. They identify the following affordances: 

\begin{itemize}
\renewcommand{\labelitemi}{$\bullet$}
\item A single sheet is light and physically flexible.
\item It is porous, which means that it is markable and that marks are fixed and spatially invariant with respect to the underlying medium.
\item It is a tangible, physical object.
\item Engagement with paper for the purpose of marking or reading is direct and local. In other word, the medium is immediately responsive to executed actions, and interaction depends on physical co-presence.
\end{itemize}

They conclude by stating that in order to drastically reduce paper usage, people need to reorganize the way they work. Specifically, they need to align themselves more with the systems they work with.

In an earlier publication \cite{Sellen:1997:PAR:258549.258780}, Sellen et al. say that there is very little research that has looked systematically role of paper in organizational life. They claim that this is partly due to paper being perceived as a symbol of the uninteresting past. Sellen et al. insist that in order to design technologies that might one day replace paper, designers need to pay attention to how paper is used and the affordances it provides. As mentioned previously, this is what our study will focus on: how paper is used and the underlying reasons behind its usage.

Plimmer and Apperley make a different argument \cite{Plimmer:2007:MPW:1278960.1278961}. They claim that long term success will depend on computer systems better supporting human systems. They recommend an activity centred model of analysis and design for technologies that are being built to support paperless environments. They go on to present Penmarked, a paperless environment for annotating and marking student assignments which they claim addresses many of the affordances of paper that are not available in standard systems. While Penmarked might make the job of a busy teaching assistant (TA) easier, it does not replicate some of the affordances of paper that make it a go-to choice for the participants I interviewed. We will discuss this point in more detail later on in this paper.

In a more recent publication \cite{Thayer:2011}, Thayer talks about some schools and colleges that are enthusiastically embracing eReaders and tablet computers with the mindset that they must update their methods and tools in order to run a classroom more efficiently. He argues that this mindset leads to the paperless school failing, in a way. By not addressing the affordances lost as a result of eliminating paper, they fail to set up a paperless system that would work in the long term. He recommends that HCI researchers need to partner with education researchers to examine the impact of switching to such devices from paper and apply current knowledge about the use of paper in workplaces to the context of schools. Once again, Thayer's findings provide further motivations for our study.

Like the aforementioned researchers, we believe that in order to design an alternative that would truly eliminate paper use, we need to not only look at what paper is used for and its affordances, but also the attributes that account for its usage and affordances.

The goal of this study is to understand what properties of devices and what properties of paper motivate our participants to use them. This will help identify the pieces that are missing in current technologies but are possessed by paper.

It's important to note that previous research done in this domain has not focussed on academia as a whole. While the research done by Plimmer et al  \cite{Plimmer:2007:MPW:1278960.1278961} looks into marking and annotation practices of TAs at universities, their participant pool does not represent other roles in higher education. Thayer's work \cite{Thayer:2011} focusses on the use of E-readers by students at schools and colleges but not teaching staff or faculty members. For their research, Sellen et al \cite{Sellen:1997:PAR:258549.258780}, \cite{sellen:myth} studied the work practices of knowledge based and document intensive businesses.

\section{METHODOLOGY}

\subsection{Recruiting}

A total of eight participants were recruited for this study, all of whom were affiliated with the University of Waterloo. Five of the eight participants were graduate students, one was a research assistant who had recently completed her master's at UW, one was a lecturer and recently completed his master's at UW, and one was a researcher and lecturer.

To minimize bias, the real purpose and goal of this study was not revealed to participants. Rather, the study was presented as one trying to better understand and analyze the work practices of people in academia.

Participants were contacted to set up an in-person interview. Prior to the interview, participants filled out a short online survey.

\subsection{Survey}

The online survey contained the following questions:

\begin{itemize}
\renewcommand{\labelitemi}{$\bullet$}
\item What portable computing devices do you use?
\item How often do you use these devices for academic work (such as assignments, research, lectures, notes, etc)?
\item How often do you use paper for academic work?
\item How often do you receive paper documents from others, for academic work?
\item Are there any other tools (such as white boards, etc) that you use frequently while doing academic work?
\end{itemize}

The purpose of this survey was to collect relevant information from the participants before the interview. Also, by asking these questions in the form of a pre-interview online survey, the participants were given the opportunity to take their time and answer these questions accurately.

The last question in the survey was included to get an idea of how often the participants end up working with paper deliberately and how often paper documents are handed to them by another party.

\subsection{Interview}

Each one-on-one in-person interview was approximately forty-five minutes long and semi-structured in nature. We met five of the participants at his/her primary workspace and three of them at his/her secondary workspace. Every interview was captured via short-hand notes and an audio recording. 

To collect some background information and get better acquainted with the participant, I began by asking each participant about to his/her role at the university, area of expertise and research and duration at the university. These findings along with some of the answers provided in the online survey are presented in Table 1.

\begin{table*}
\centering
\caption{Participants' roles, devices and frequency of Device and paper use.}
\begin{tabular}{c|c|c|c} 
P no.& Devices  & Device Usage & Paper Usage\\ \hline
P1 & Laptop & Few time a week & Everyday\\ \hline
P2 & Tablet PC & Few times a week & Everyday\\ \hline
P3 & Laptop, Tablet, E-Reader & Used in business& Everyday\\ \hline
P4 & Laptop & Unexplained usage& Everyday\\ \hline
P5 & Laptop, Tablet, E-Reader & Few times a month & Most days of the week\\ \hline
P6 & Laptop & Everyday & Never\\ \hline
P7 & Laptop & Everyday & Everyday\\ \hline
P8 & Laptop & Everyday & Everyday\\ \hline
\end{tabular}
\end{table*}

The interview began with broad, open ended questions that referred to some of the answers provided by the participant in the online survey. For example, P1 was asked ``You stated in the online survey that you use your laptop a few times a week. Could you tell me about what kind of tasks you use your laptop for?" Depending on the answers provided by the participant, we asked more specific follow-up questions were asked. For example, P4 said that he preferred to use his laptop to skim over articles and papers quickly but printed out documents that he needed to ready thoroughly. I followed up by asking him ``Why do you think you prefer to use your laptop to skim but prefer to print when reading something in detail?" All the answers provided in the online survey were followed up in this manner. 

Other questions posed during the interview included:

\begin{itemize}
\renewcommand{\labelitemi}{$\bullet$}
\item Can you tell me a little bit about how your workspace is laid out? 
\item What are your work practices like when you are travelling?
\item Have you ever considered purchasing any of the portable computing devices available in the market today to help you with academic work? 
\end{itemize}

\begin{itemize}
\renewcommand{\labelitemi}{$\bullet$}
\item Can you tell me a little bit about how your workspace is laid out? 
\item What are your work practices like when you are travelling?
\item Have you ever considered purchasing any of the portable computing devices available in the market today to help you with academic work? 
\end{itemize}

\subsection{Analysis}

The audio recordings of the interviews were transcribed and the data set was analyzed using an affinity diagram. Participant quotes that expressed how they used a certain medium, their motivations behind it, and other relevant information were extracted from the data set. These quotes were then grouped by ``How medium is used" and ``Attribute of medium" to identify broader themes.

\section{FINDINGS}

Some strong common themes arose from the analysis of the dataset along with a few outliers which were the result of the unique personal preferences and working styles of some participants.

\subsection{Paper}

With the exception of P6, all the participants used paper quite frequently for academic work.

\subsubsection{Arrangement, Navigation and Annotation}

When it came to reading lengthy documents (typically, academic papers) in detail, the seven participants said that they preferred to print them than read it on a device screen. When asked why, six of these participants stated that printed copies allowed them to view more information at a time. P5 said,  \emph{``The layout of information and spatial arrangement is very important to me. I can lay out the pages so that I'm looking at an entire article at once and I know where a certain paragraph is in relation to another."}

A similar comment was made by P1: \emph{``A lot of times I'll be working with multiple papers. I like to spread all the pages out on my desk so I can see all the pages. If I had to scroll up and down to find something I had seen on another page, it would disrupt my work flow. I get distracted easily."}

P2 stated that he liked how easily he could flip through pages and refer back to previous sections in a document \emph{``without having to scroll, which gets very annoying."}

Five participants mentioned that printed copies of documents allowed them to mark up and edit more easily.

P3 said, \emph{``I could open up the PDF document on my computer and mark it up on Acrobat but other than being just too cumbersome, it messes with your flow of thought."}

P1 stated that she caught \emph{``more mistakes on paper than on a PDF."}

\subsubsection{Easy To Read}

Four of the seven participants cited eye strain and readability as one of the reasons they preferred reading content heavy documents on paper.

P7 said that he found reading something on paper \emph{``easier on the eyes".} He added that with paper he didn't have any \emph{``lighting or glare issues."}

On a slightly different note, P4 said\emph{``Apart from the light of the screen shining on my face, I can hold paper however I want while reading it. It seems unnatural to have something upright and in front of you for reading. I prefer to have it laid horizontally on a surface. I guess you could do that with a tablet or something but I don't have one."}

\subsubsection{Open Format}

All seven participants preferred to use paper when brainstorming, taking unstructured notes or doing rough work. When I asked why paper was a better tool in such instances, P4 said, \emph{``Paper has an open format. You can cross things out easily, and write in all kinds of notations. Pen and paper just respond better when you're in the moment and really want to get ideas flowing. Typing something out just doesn't have the same flow."}

P4 went to on to say \emph{``Paper is very familiar and I think that also has something to do with it being the right tool for brainstorming."} P1 mirrored this sentiment and said that she thought that the \emph{``familiar and comforting quality"} of paper encouraged her flow of ideas. 

P5 said that paper makes it \emph{``very easy to put things down in a way that seems logical at that time."} He added, \emph{``You can draw arrows, boxes, whatever you want, to clearly articulate that idea."}
P2 mentioned that he had taken free hand notes on his tablet PC in lectures in the past but hadn't used it as much while working on his PhD research. He said \emph{``I would take it to class with me and jot down notes for the lecture but I haven't used that feature since I finished my coursework."} I followed up by asking him why he didn't use his tablet PC's note taking feature anymore and learnt that he did most of his post-coursework research on his desktop and didn't want to have another computer around just for notes. \emph{``I don't have a very good reason but I think paper is just more convenient when I'm already working on my desktop"} P2 said.

P8, who had used a tablet PC in the past, had a similar comment.\emph{"When I had it, I found it to be decent for taking notes in lectures but it wasn't my tablet PC. I had only been borrowing for a few months and I went back to paper once I returned it to the owner"} he said.

\subsubsection{Organization}

Next, I asked the participants how they managed the papers that they worked with. Interestingly enough, all seven said that they didn't have a very good way of dealing with and organizing the all the paper documents they worked with.

Four of the seven participants said they typically keep their printed documents in piles and stacks on their desk but that it isn't very effective way to organized them long term.

P1 said \emph{``I'll print anything that I think I might want to read at some point so I end up with quite a few papers on my desk,"} she said. \emph{``The stacks of paper kind of act as a gentle reminder that I need to read them at some point, that I have work to do,"} she added. \emph{``But sometimes they get mixed up and shuffled around and I won't be able to find a particular paper I am looking for."} This leads to her reprinting the documents she cannot locate.

According to P3, organizing her notes had become unmanageable to the point where she had to come up with a better archival system. \emph{``I write on my note book and then take a picture the page with my iPad. After that I tag that image with relevant tags. So now I can search and keep track of my notes electronically and I'll never have to go back to the actual piece of paper that I originally scribbled on,"} she said. \emph{``Up until I got the iPad, I would typeset my notes on LaTeX but that just wasn't convenient enough for me to keep up with."}

P7 stated that he tries not to hold onto every document that he had ever printed but only those he deemed \emph{``important enough".} He said that he organizes his papers by filing them in hanging folders in his desk, a practice he picked up from former colleagues. He added that this method of organization wasn't necessarily easy to keep up with. \emph{``At some point though, that line of ‘important enough' begins to blur and it becomes very easy to hoard. I think I have been doing a good job so far of not cluttering my workspace with a gazillion printed articles."}

P8 said that he archives his printed documents \emph{``just to be safe"} but disposes of them after at a later time when he feels that he doesn't need them on hand anymore. 

\subsection{Devices}

P6 was the only participant I interviewed that almost never uses paper and relies on his laptop and desktop for all academic tasks. He attributes part of this working style to his programming background. \emph{``I've been programming for a long time and as a result have gotten very comfortable with looking at things on screen,"} he said. P6 went on to say that he finds reading on paper uncomfortable and needs to wear to reading glasses when doing so. He added, \emph{``I find it hard to decipher handwritings so that's another reason I like to read on the monitor."}

When I asked him about his brainstorming practices, P6 said that he often starts with\emph{``more concrete ideas than abstract ones"} and prefers to\emph{ ``start typing things out on the computer right away."} He added \emph{``I realize I might be a bit peculiar and this is not how most other people work but it's how I think and work."} 

Devices are the tools of choice for all participants when it came to tasks that involved programming and performing complex computations. They are also crucial for a variety of other academic activities.  On closer inspection of these activities, I found that the following attributes of devices led to them being used the way they were.

\subsubsection{Portability and Storage}

The participants that do travel often for academic purposes said that they rely on their portable computing devices on their trips.  

While there are some concerns about losing these devices on such trips, the participants said they would rather carry a few essential items with them instead of binder or stacks of paper. 

P4 said \emph{``If I'm travelling, I only take my laptop with if I'm going to be presenting something, because I'm nervous about losing it. Otherwise, I'll only take the papers that I absolutely know I'll need."}

When travelling, the participants were willing to compromise and part with some of the affordances that paper provided for the sake of a more convenient and light trip. P1 said \emph{``I would really like to read on my laptop when I travel so that I don't have to take any paper with me. My current laptop is on its last leg though and keeps dying on me so until I get a new one, I'll have to keep travelling with a bunch of binders."}

P3, who doesn't otherwise use her eReader said that it's her preferred tool when travelling. She said \emph{``If I have a bunch of papers when I travel I feel like I might misplace them so I'll read the PDFs on my Kindle. The Kindle isn't cheap and I invested money into it so I'll definitely be more careful about not losing something like that."} 

The participants that use paper on regular basis also keep electronic copies of documents. P3 explained \emph{``If I have the PDF, I know I have that document to refer to at a later time. Whereas if I just have a printed copy, who knows where it will end up once I'm done with. It's much easier to store and keep track of things electronically."} P1 said \emph{``If I have a document in my Dropbox\footnote{Dropbox is a web-based file hosting service operated by Dropbox, Inc. Please visit www.dropbox.com for more information. }, I can access it from anywhere. I was recently in Toronto for a conference and was mostly using my boyfriend's laptop and I could download what I needed from my Dropbox. But with a paper copy, if I don't have it on me, there's not much I can do about it."}

The physical space required to store paper artefacts was an issue mentioned by P6. This participant has a large library of physical books at his primary workspace but the electronic library on his computer is even larger. \emph{``I have many more eBooks on my computer. They're all organized and in folders and much easier to store. I can't remember the last time I searched a physical book for something and now they're quite out of date."}

\subsubsection{Sharing and Searching}

When it comes to sharing documents, the participants prefer to do so electronically. A couple of participants, however, have received printed copies of documents that they were asked to proofread. P1 said \emph{``I'll get printed copies from someone if I'm proofreading a document for them. I'm kind of doing them a favour and they're aware of that so they'll be polite and give me a paper copy because that's easier to read. I think I'd do the same if I were to ask someone to proofread something I had written."}

Even though the participants preferred to read lengthy documents on paper, they chose to skim articles on their devices. They claimed that this helped them identify articles that didn't require a detailed read and thereby, minimized unnecessary printing. The ability to search electronic documents seems to support this kind of behaviour as well. P2 said \emph{``I try not to print anything I don't have to read thoroughly. I can look at it quickly on my computer and press control F to find a specific word or phrase. Why waste paper then."}

A similar comment was made by P5:

\emph{``If I know I'm looking for something particular in an article, it's quite easy to search and find it. You can't say the same about finding things in printed documents though."} 

P6 claimed that search was one of the reasons he had moved away from using paper. \emph{``You can't search books or paper. There is really no easy way of doing that even with indices. If I'm looking for something specific, only searching by term will allow me to find it, and I can do that best on the computer."} 

\subsubsection{Formal Writing}

All eight participants said they preferred to use their devices when writing up anything formally.

P4 said \emph{``When I start formally writing something, I don't necessarily do it in a linear fashion. I might write point two before I write point one but something about typing it out in a word processor helps me maintain the structure of the article." } 

According to P2, he types up anything that he thinks might be shared with others at some point. \emph{``I type up my handwritten notes that I end up sharing with my supervisor or anyone else. Basically anything that I want to write in a structured manner, I will type up on my computer."}

\subsection{Other Tools}

\subsubsection{Whiteboards}

P2, P3, P7 and P8 stated that they like to use whiteboards when brainstorming. These participants claimed that whiteboards offered them the same affordances as paper but on a much larger scale. P3 said \emph{``The whiteboard works as well as paper, but you just have more real estate".}

According to P8, \emph{``Having a marker in one hand and an eraser in the other just helps facilitate the process a little bit more."}

P1 and P4 said they liked to use whiteboard when working with other people. P1 explained, \emph{``If I'm working with other people, I guess we could sit around a desk and work on a piece of paper but a whiteboard is just more convenient for something like that. I'd use paper if it was just me working."}

\section{Discussion}

As pointed out by prior research in this domain, it is the inherent affordances of paper that make it such a powerful tool for humans. The findings of this study confirm that. We will take a closer look at these findings in this section.

One of the main motivations behind paper use, as cited by the participants, is the fact that it can be arranged in numerous ways, spread out on surfaces and flipped through. These qualities allow the reader to get an overview of the layout and content in a fairly short amount of time by providing spatial and tactile clues as markers to particular parts of the documents, as mentioned by Plimmer et al \cite{Plimmer:2007:MPW:1278960.1278961}.  None of these qualities of paper have been successfully emulated by any of today's technologies. One might argue that with multiple devices and screens, we can tackle the issue of ``viewing more" at a time. While that might be an acceptable solution to some users, albeit an expensive one, it only addresses one piece of the puzzle and does not come close to replicating the other conveniences of paper. 

The above mentioned qualities are also responsible for paper being the preferred medium for active reading \cite{Plimmer:2007:MPW:1278960.1278961}. Depending on the level, active reading can include activities from scanning documents quickly to detailed word-by-word level scrutiny. Annotating, underlining and highlighting are encouraged for active reading as it helps the people emphasize and remember information. Given these facts, it's not surprising that the participants prefer marking up and editing printer versions of articles to their electronic counterparts. Other than keeping the readers focussed and not disrupting their flow of work, active reading on paper helps one spot more mistake in a shorter period of time, as mentioned earlier by P1. There is a very clear gap between this kind of experience and marking up a PDF document on- screen. Not only would electronic marking up demand more from the reader, cognitively, more steps are needed to type (or write, if using a pen-based system) up the comment itself.

Reading can be a visually demanding activity, especially if the document being read is a long and complex one. Other than some eReaders that use e-ink in their displays, majority of devices use backlight illumination. Staring at a backlit screen can cause eye strain and even headaches after prolonged viewing. While E-ink based eReaders might be pleasant to read from, reading PDFs on such devices requires quite a bit of scrolling and adjusting zoom settings. 

Another key attribute of paper that lends to its popularity as a medium is its forgiving and open format. One can write, draw and use complex notations on paper without having to worry about lags, proper handwriting display, touch and stylus detection, etc. P4's quote about pen and paper responding immediately and flawlessly to the author's hand movements and whims sums up why paper works so well for brainstorming. Whiteboards are popular because of their open format as well but, as our participants mention, it is perhaps their size and reusability make them great tools for group work and ideation. However, their large size is what makes whiteboards difficult to carry and move around which in turn limits their use. 

A downside of paper is that it requires physical space for storage. This might not be an issue when dealing with small amounts of paper but can be quite the challenge when it comes to managing larger volumes. Most of the participants I spoke to prefer to hold on to printed copies of documents for some period of time and as a result have to deal with organizing and storing them in some way. Having worked with paper for years, each participant has devised a way to manage paper in a way that works for him/her most of the time. They admit, however, that they do not have an elegant, quick and convenient organization and archival system in place that would work on a long term basis.

Electronic storage and organization of documents, on the other hand, isn't as challenging. Physical space is no longer a huge constraint since one can store as many documents on a device as its hard drive space will allow. This is also the reason why devices can be more portable than paper and easier to carry when travelling. 

There are also many online tools available today (for example, Dropbox) that make electronic storage even more convenient. What makes these types of applications popular is the fact that users can access their account from any computing device with an internet connection and upload and download any file to and from it as they like.

The biggest advantage of electronic archives is perhaps the ability to easily search them. While indices, tabs and table of contents in books and paper documents can help a user locate a specific section or word, they don't match the convenience and speed of electronic search. Based on some of the statements made by the participants, search is a powerful enough feature to trump some of the inconveniences of on-screen reading. 

Sharing electronic documents is also more convenient than sharing paper documents. Paper documents might get lost in the mail, destroyed, stolen, and can be downright expensive to ship around the world. None of these issues exist with electronic sharing though. While email might still be the preferred way of sharing electronic files, applications such as Google Docs are making it easier for people to collaborate remotely as well. 

It's quite clear that paper and devices have their pros and cons and with the exception of P6, all the participants use both media for academic work. This is not surprising given the current state of technology – some things can be done better on devices but for other tasks, there is no good alternative for paper. However, there seems to be a general belief that the emergence of powerful portable computing devices will replace paper. While it might be a small step in that direction, technology has to progress a fair bit before it can eliminate paper use from the lives of humans. In their publications, Sellen et al. imply that until a suitable alternative for paper is available in the market, paper usage might surge with the growth of technology. This implication rings true in P3's final comment during my interview with her, which was: \emph{``To summarize, I would say technology hasn't reduced my paper usage. If anything, it is enabling me to use paper."}

\section{CONCLUSIONS AND FUTURE WORK}

This preliminary study has confirmed the findings and recommendations of researchers who have looked into the viability of a paperless world and shown that this is indeed an interesting area of study. 
Apparently unassuming, paper is a very powerful tool, thanks to its intrinsic attributes and affordances. It has been used for over a thousand years and has heavily influenced the progress of many different fields. While paper is easy to use and feels rather natural to work with for most human beings, it does have some disadvantages. Storage, information retrieval, preservation, loss and damage, are some of the issues, to name a few, that people have to deal with in this territory. Perhaps one of paper's biggest downfalls is its environmental impact. There have been recent attempts to create alternatives that are more eco-friendly. Unfortunately, these alternatives do not address all the issues we face with paper today and apart from being expensive, are not readily available to the public.

Looking at the data collected, we see that our participants value paper because of its open format, tangibility as a physical object, readability, and easy navigation. These attributes and affordances make paper a rather spontaneous tool that is easy to annotate and brainstorm with while acting as a reminder to do something. The familiarity of paper, as put by one of our participants, lets its user work in a rather smooth way without disrupting the flow of the task at hand. While this study was rather small and we have to be careful drawing conclusions from it, there are parallels between the observations of experts and researchers in this field and those made here. 

We believe the next step in furthering the research done here would be to conduct some larger and more in-depth studies that delve deeper into the attributes of paper and how they can be mimicked by devices. Quantitative task-based comparative studies in which users perform similar tasks on paper and devices would also provide interesting insights. As mentioned previously, there is an opportunity to leverage these findings to shape technology and design devices that offer people similar affordances to paper.

A successful alternative to paper will have to emulate its positive attributes while addressing some of its less desirable properties in a sophisticated manner. It is possible that we do not currently possess the necessarily technologies to devise such an alternative at this time. With continued in-depth research and scientific advancements, however, we do believe that we can, one day,  ``go paperless." 

\balancecolumns

%
\bibliographystyle{abbrv}
\bibliography{jc-report}  
%
%

\balancecolumns

\end{document}